\documentclass[aps,prc,twocolumn,floatfix,showpacs,amsmath,amssymb,superscriptaddress]{revtex4}
\usepackage{graphicx}
\usepackage{dcolumn}% Align table columns on decimal point
\usepackage[usenames]{color} % To use colored text to denote comments
  \def\nuc#1#2{\relax\ifmmode{}^{#1}{\protect\text{#2}}\else${}^{#1}$#2\fi}
  \def\itnuc#1#2{\setbox\@tempboxa=\hbox{\scriptsize\it #1}
    \def\@tempa{{}^{\box\@tempboxa}\!\protect\text{\it #2}}\relax
    \ifmmode \@tempa \else $\@tempa$\fi}
  \newcommand{\beq}{\begin{equation}}
  \newcommand{\eeq}{\end{equation}}
  \newcommand{\bea}{\begin{eqnarray}}
  \newcommand{\eea}{\end{eqnarray}}
  
  \newcommand{\eq}[1]{Eq.~\eqref{#1}}
  \renewcommand{\sec}[1]{Sec.~\ref{#1}}
  \newcommand{\fig}[1]{Fig.~\ref{#1}}
  \newcommand{\co}{(Color online)}
  \newcommand{\hf}{Hauser-Feshbach}
  \newcommand{\we}{Weisskopf-Ewing}
  \newcommand{\cn}{compound nucleus}
  \newcommand{\rmref}{\ensuremath{\mathrm{ref}}}
  \renewcommand{\ng}{\ensuremath{(n,\gamma)}}
  \newcommand{\zrzero}{\nuc{90}{Zr}}
  \newcommand{\zrone}{\nuc{91}{Zr}}
  \newcommand{\zrtwo}{\nuc{92}{Zr}}
  \newcommand{\zrthree}{\nuc{93}{Zr}}
  \newcommand{\zrfour}{\nuc{94}{Zr}}
  \newcommand{\zrfive}{\nuc{95}{Zr}}
  \newcommand{\zrsix}{\nuc{96}{Zr}}
  
\begin{document}
\preprint{UCRL-JRNL-XXXXXX}

\title{Determining neutron-capture cross sections via the surrogate
  reaction technique} 
\author{C. Forss\'en}
\email[]{c.forssen@fy.chalmers.se}
\affiliation{Fundamental Physics, Chalmers University of Technology,
  SE-412 96 G\"oteborg, Sweden}
\affiliation{Lawrence Livermore National Laboratory, P.O. Box 808, L-414, 
Livermore, CA  94551, USA}
\author{F.~S.~Dietrich}
\author{J.~Escher}
\author{R.~D.~Hoffman}
\affiliation{Lawrence Livermore National Laboratory, P.O. Box 808, L-414, 
Livermore, CA  94551, USA}
\author{K.~Kelley}
\altaffiliation[Present address: ]{Department of Physics, Brigham Young
  University-Idaho, 525 South Center Street, Rexburg, ID 83460, USA}
\affiliation{Lawrence Livermore National Laboratory, P.O. Box 808, L-414, 
Livermore, CA  94551, USA}

\date{\today}

\begin{abstract}
Indirect methods play an important role in the determination of nuclear
reaction cross sections that are hard to measure directly. In this paper
we investigate the feasibility of using the so-called surrogate method
to extract neutron-capture cross sections for low energy
compound-nuclear reactions in spherical and near-spherical nuclei. We
present the surrogate method and develop a statistical nuclear-reaction
simulation to explore different approaches to utilize surrogate reaction
data. We assess the success of each approach by comparing the extracted
cross sections with a predetermined benchmark. In particular, we employ
regional systematics of nuclear properties in the $34 \leq Z \leq 46$
region to calculate \ng\ cross sections for a series of Zr isotopes, and
to simulate a surrogate experiment and the extraction of the desired
cross section. We identify one particular approach that may provide very
useful estimates of the cross section, and we discuss some of the
limitations of the method. General recommendations for future
(surrogate) experiments are also given.\\
\end{abstract}
% insert suggested PACS numbers in braces on next line
\maketitle
%
%**************************************************************************
\section{\label{sec:intro}Introduction}
Nuclear reaction cross sections are often difficult to measure
directly. This is particularly true for reactions relevant to
applications in nuclear astrophysics, since radioactive nuclei play an
influential role in many cosmic phenomena but cannot be easily studied
in the laboratory. Information on these nuclei, and on their relevant
cross sections, is needed to improve our understanding of the processes
that shape our universe. While indirect methods for determining
direct-reaction cross sections have become very popular in recent years
\cite{smi01:51}, compound-nuclear reaction cross sections are typically
determined purely theoretically.

For nuclei with mass $A \geq 30$, statistical-reaction model
(Hauser-Feshbach) calculations are widely used to estimate cross
sections that have not been measured. These calculations require input
data such as masses, one- and two-particle separation energies,
properties of resonances, level densities, optical potentials for
particle transmission coefficients, and gamma strength functions. This
data should be constrained by measurements wherever possible, but for
the thousands of nuclei involved in astrophysical environments one has
to rely on global phenomenology or, alternatively, on microscopic
nuclear theories.

In this paper, we explore the possibility of using an indirect method,
the surrogate nuclear reaction method, for obtaining \ng\
compound-nuclear reaction cross sections.  A simplified version of the
method, which combines experiment with reaction theory to obtain cross
sections for reactions that proceed through a compound nucleus, was
first used in the 1970s~\cite{cra70:41,bri79:72} to extract ($n,f$)
cross sections for various actinides from transfer reactions with $t$
and \nuc{3}{He} projectiles on neighboring nuclei, followed by
fission. A modern version of the approach was used by Petit \emph{et
al.} to study the \nuc{233}{Pa}$(n,f)$ reaction cross section using a
$(\nuc{3}{He},p)$ transfer reaction~\cite{pet04:735}. More recently,
Burke~\emph{et al.}~\cite{bur06:73} and Plettner~\emph{et
al.}~\cite{ple05:71} constructed ratios of decay data from surrogate
experiments on two different uranium isotopes, and used that information
to extract the \nuc{237}{U}$(n,f)$ cross section. The surrogate ratio
method, as applied to actinide nuclei, was examined in much detail by
Escher and Dietrich~\cite{esc06:74}. These efforts have shown that the
surrogate method is a very useful tool for predicting various $(n,f)$
cross sections in the actinide region, and there is no \emph{a priori}
reason why the method should be limited to studies of fission. In
principle, the surrogate method can be applied to any reaction that
proceeds via a well-defined, equilibrated compound state; but its
greatest potential value lies in applications that involve unstable
isotopes. Among the unanswered mysteries about the nature and evolution
of our universe is the origin of the heavy elements. Much effort is
currently being devoted to exploring possible paths for the
nucleosynthesis (such as the s and r processes) and astrophysical
environments that can produce the elements between iron and uranium. Of
particular interest in the context of the s process are \ng\ reactions
on branch point nuclei, unstable isotopes with a life time long enough
to allow the reaction path to proceed by either neutron capture or
$\beta$ decay. In principle, these isotopes are ideally suited for
investigations using the surrogate method since they are located next to
stable elements that can be used as targets in the surrogate
experiment. Surrogate approaches for other neutron-induced reactions
have been considered as well. Early experiments~\cite{bri76:unp} were
carried out to assess the feasibility of using the surrogate technique
to determine cross sections for ($n,\alpha$) and ($n, p$) reactions on
nuclei in the mass-90 region. These experiments highlighted several
issues that needed to be addressed in order to extract reliable cross
sections from surrogate measurements. 

The experiments mentioned above were analyzed under the simplifying
assumption that the decay probabilities are independent of the
particular spins and parities of the compound-nuclear states that are
occupied in the neutron-induced as well as in the surrogate
reaction. This assumption, which is known as the Weisskopf-Ewing
limit~\cite{wei40:57,gad92:book}, is not always valid and its
application needs further exploration. We will investigate the validity
of the \we\ approximation for \ng\ reactions involving spherical, or
near-spherical, targets in the Zr region.  The fact that the spin-parity
($J\Pi$) distributions in the decaying compound nucleus can be very
different in the $n$-induced and the surrogate reaction is referred to
as the ``$J\Pi$ population mismatch''. For s-process branch points,
e.g., low-energy neutrons bring in very little angular momentum, while
direct reactions leading to the same \cn\ can produce very different
angular-momentum distributions. This leads to challenges for extracting
information from a surrogate experiment, as was already recognized from
the early experiments in the mass-90 region~\cite{bri76:unp}.

Recently, Younes and Britt~\cite{you03:67,you03:68} demonstrated that
taking into account the $J\Pi$ population mismatch can have a
significant effect on the extracted results.  They revisited the data
from the original surrogate transfer-reaction induced fission
measurements from the 1970s, and employed a simple direct-reaction model
to account for the angular-momentum population difference between the
neutron-induced and direct reactions. They were able to improve on
earlier results for ($n, f$) cross sections for various Th, U, Np, Pu,
and Am isotopes. For \nuc{235}{U}, Younes and Britt reproduced the known
fission cross section for the $J = 7/2^+$ ground state and were able to
predict the fission cross section for the isomeric $1/2^+$ state at
77~eV, which to date has not been measured directly.

In this paper we investigate the feasibility of applying the surrogate
method to extract low-energy \ng\ cross sections on mass $\sim 90 - 100$
nuclei. There are essentially two different sources of uncertainty
inherent in the surrogate method: (i) Insufficient knowledge of the
decay pattern for the relevant compound nucleus, which must be
supplemented by reaction modeling; (ii) Insufficient knowledge of the
spin-parity distribution of the decaying compound nucleus.  We will
present calculations that illustrate the effect of these two sources of
uncertainty on cross sections extracted from surrogate experiments.  We
will consider and compare different strategies of utilizing the
surrogate data for obtaining unknown compound-nuclear reaction cross
sections.

We will present calculations for a range of zirconium isotopes and in
particular we perform surrogate experiment simulations to study the
extraction of the \zrone\ng\zrtwo\ cross section. However, our
discussions are more general and can in principle be applied to most
spherical and near-spherical nuclei in the mass 90-100 region. This mass
region is not only of importance for understanding the s-process
nucleosynthesis path, but also encompasses the majority of light fission
fragment nuclei of the binary fission yield distribution. Consequently,
there is great interest in obtaining cross sections for neutron-induced
reactions involving nuclei in this region.
A particularly interesting application appears in nuclear astrophysics:
\zrthree\ and \zrfive\ are both s-process branch points and are believed
to be produced in AGB stars. It has been proposed~\cite{her05:43} that
the relative abundance of \zrfour\ and \zrsix, as measured in presolar
grains~\cite{nic97:277}, depends on the efficiency of mixing between the
H- and He-burning shells of AGB stars. Thus, better knowledge of the
neutron capture cross sections of \zrthree\ and \zrfive\ can lead to a
diagnostic tool to probe stellar interior physics.
%
%**************************************************************************
\section{\label{sec:method}Method of the Study}
In the present work we use a statistical-reaction model simulation to
assess whether the surrogate method can be employed to extract
low-energy \ng\ cross sections.  First, we demonstrate how the typical
level of uncertainty in the \hf\ parameters affects the cross sections
obtained in a purely theoretical approach.  The cross section that best
represents the available data will later on serve as a benchmark for the
surrogate method.  The combination of models and parameters used to
produce this benchmark cross section is referred to as the ``Reference
Decay Model'' and will be considered as the most realistic description
of the true decay of the compound nucleus.  We simulate the impact of
having insufficient information about the ``true'' decay by considering
variations of the decay model.  We also use the same set of decay models
when studying the decay of the compound nucleus populated through a
surrogate reaction.  This type of theoretical approach has a number of
distinct benefits:
(i) We are able to access physical quantities that are not directly
  observable in an experiment, such as spin-parity dependent branching ratios
  for individual exit channels.
(ii) We can alter key quantities such as level densities and gamma
  strength functions and carry out sensitivity studies.
(iii) Performing simulations will allow us to identify the main
  limitations of the method and to quantify the precision that can be
  expected.

All calculations are carried out with a modified version of the \hf\
code \textsc{Stapre}~\cite{uhl76:irk,str80:ictp}. In order to account
for the fact that a direct reaction produces a $J\Pi$ distribution in the
residual nucleus that is different from the one associated with the
desired neutron-induced reaction, we have modified the code so that it
is suitable for surrogate-reaction studies. In particular, we included
an option to allow the $J\Pi$ distribution of the first \cn\ to be read
in from a file rather than calculated from entrance-channel transmission
coefficients. It is therefore possible to specify an arbitrary $J\Pi$
distribution for a given \cn\ and to predict the decay of the
nucleus. In addition, we have implemented the specific models for level
densities and photon strength functions that we are using for this work
(see Sec.~\ref{subsec:decmod}).

The theoretical framework for the surrogate method is described in
\sec{sec:theory}. Statistical-reaction calculations are performed using
recently developed regional systematics for nuclear properties in the
$34 \leq Z \leq 46$ region. These systematics, and the resulting
estimated \ng\ cross sections for a range of zirconium isotopes, are
presented in \sec{sec:zr}. In \sec{sec:surrogate} we use the reference
decay model obtained from the regional systematics to simulate a
surrogate experiment. We discuss three different approaches for
utilizing surrogate data to extract low-energy neutron capture cross
sections. We perform sensitivity studies and discuss theoretical and
experimental challenges for the surrogate method. Concluding remarks and
recommendations are given in \sec{sec:conc}.
%
%************************************************************************** 
\section{Statistical reaction theory and the surrogate idea%
\label{sec:theory}}
\subsection{\hf\ theory}
The formalism appropriate for describing compound-nucleus reactions is
based on the statistical \hf\ theory~\cite{hau52:87,gad92:book}. In this
section we summarize the \hf\ theory in a form suitable for application
to surrogate experiments that will be discussed in
\sec{sec:surrtheory}. Let us consider the case of a reaction leading
from an entrance channel $a + A$ (denoted $\alpha$) via an intermediate
compound-nuclear state $B^{*}$ to an exit channel $c + C$ (denoted
$\chi$). The energy-averaged cross section for this reaction, as a
function of the center-of-mass energy $\varepsilon_\alpha$ in the
incoming channel, is written as
\beq
    \sigma_{\alpha\chi}(\varepsilon_\alpha)
    = \sum_{J,\Pi}
    \sigma_{\alpha}^{}(\varepsilon_\alpha,J,\Pi)
    {G_{\chi}^{}(U,J,\Pi)},
\label{eq:desired}
\eeq
where we have assumed that the Bohr hypothesis of independence between
formation and decay holds separately for each value of total angular
momentum and parity, $J\Pi$, of the \cn. The formation cross section, 
$\sigma_{\alpha}^{}(\varepsilon_\alpha,J,\Pi)$, for particular values of 
$J\Pi$ in the \cn, as well as the branching ratio of decay into
channel $\chi$,  $G_{\chi}^{}(U,J,\Pi)$, are both energy-averaged
quantities. The excitation energy $U$ of the \cn\ is related to the
separation energy $S_a$ of particle $a$ by $U = S_a +
\varepsilon_\alpha$.

For radiative capture reactions at low energies one is usually
interested in the cross section for producing a particular
isotope. Therefore, we will define $G_{\chi}^{}(U,J,\Pi)$ as the
branching ratio of decay into channel $\chi$ integrated over all bound
states of the residual nucleus
\begin{widetext}
\beq
  G_{\chi}^{}(U,J,\Pi) = 
  \frac{\sum_{l_\chi,s_\chi}\sum_{J_C,\Pi_C} 
  \int_0^{U-S_c} dU_C
  T_{\chi}(\varepsilon_\chi,l_\chi,) \rho_C(U_C,J_C,\Pi_C)}
  {\sum_{\chi'} \sum_{l_{\chi'},s_{\chi'}}\sum_{J_{C'},\Pi_{C'}}
  \int_0^{U-S_{c'}} dU_{C'}
  T_{\chi'}(\varepsilon_{\chi'},l_{\chi'})
  \rho_{C'}(U_{C'},J_{C'},\Pi_{C'})},  
\label{eq:g}
\eeq
\end{widetext}
where $s_\chi$ and $l_\chi$ are, respectively, the exit channel spin and
the orbital angular momentum between the residual nucleus and the
ejectile. $J_C$ and $\Pi_C$ denotes the spin and parity, respectively,
of a state in the residual nucleus $C$. The excitation energy is given
by $U_C$, and energy conservation gives $U_C = U - S_c -
\varepsilon_\chi$. The terms in the sums are furthermore restricted by
parity conservation and spin-coupling conditions. The particle
transmission coefficients, $T_{\chi}(\varepsilon_\chi,l_\chi)$, can in
general depend on the complete set of quantum numbers that define a
channel, but the dependence is here limited to energy and orbital
angular momentum. For gamma-decay channels, the sum over $l_\chi,
s_\chi$ is replaced by a sum over electromagnetic multipoles, and the
transmission coefficients accordingly refer to these
multipoles. $\rho_C(U_C,J_C,\Pi_C)$ denotes the level density of the
residual nucleus. The quantum numbers and quantities for all open
channels, $\chi'$, appearing in the denominator, are defined
analogously. Finally, although the above formula does not indicate it in
the interest of simplicity, \textsc{Stapre} actually treats each step
of the decay completely and correctly; in particular, for primary gammas
populating levels that are still above the particle thresholds,
\textsc{Stapre} calculates the competition between particle and gamma
emission in dealing with further steps in the cascade.

The assumption of full independence between formation and decay of the
\cn\ can be relaxed by the introduction of width fluctuation corrections
$w_{\alpha\chi}(\varepsilon_\alpha,J,\Pi)$. These aim to correct for
correlations between the widths of the incoming and outgoing
channels. In general, the width fluctuations will only be important for
very small energies, where the effect will enhance the elastic channel
and simultaneously deplete all other channels due to flux
conservation. We will demonstrate in \sec{sec:zr} that, for the cases of
interest here, the depletion of the \ng\ cross section is $\lesssim
10$\% below 1~MeV, and negligible above that.
\subsection{The surrogate method%
\label{sec:surrtheory}}
Here we outline how the cross section of \eq{eq:desired} can be
determined by a combination of theory and experiment in the surrogate
method. In a surrogate experiment, the compound-nuclear state $B^{*}$ is
produced via an alternative (``surrogate'') direct reaction $d + D
\rightarrow b + B^{*}$, and the decay of $B^{*}$ into the desired
channel $\chi$ is observed in coincidence with the outgoing particle
$b$. In the following we assume that the ejectile, $b$, is observed at a
particular angle $\Omega_\beta$ and for simplicity we suppress the
angular dependence of the observed quantities.  The energy-differential
cross section for the direct reaction, constituting the first step of
the surrogate reaction sequence, is then
\beq
    \frac{d\sigma_{\delta\beta}}{ d \varepsilon_\beta}
    (\varepsilon_\beta) =
    \sum_{J,\Pi} \frac{d\sigma_{\delta\beta}}{d \varepsilon_\beta}
    (\varepsilon_\beta,J,\Pi),
\eeq
where $\varepsilon_\beta$ is the center-of-mass energy in the outgoing
channel $b+B^{*}$ (denoted $\beta$), and $J \Pi$ on the right-hand side
denote different spin-parity combinations of the states of the \cn\
populated via the incoming channel $\delta$ of the surrogate reaction.
This decomposition must be provided by a reaction-model calculation.
Making the non-trivial assumption that $B^{*}$ damps into a fully
equilibrated state, we can define the probability for producing the \cn\
at excitation energy $U$ with spin and parity $J \Pi$
\begin{equation}
    F_{\delta\beta}^{}(U,J,\Pi) =
    \frac{\frac{d\sigma_{\delta\beta}}{ d \varepsilon_\beta}
    (\varepsilon_\beta,J,\Pi)}
    {\sum_{J',\Pi'} \frac{d\sigma_{\delta\beta}}{d \varepsilon_\beta}
    (\varepsilon_\beta,J',\Pi')}.
    \label{eq:surrprob}
\end{equation}
For a fixed ejectile angle the reaction kinematics will give a
straightforward relation between the projectile and ejectile energies,
and the excitation energy $U$ of the \cn.

In a surrogate experiment the ejectile $b$ is measured in coincidence
with an appropriate observable that tags the channel $\chi$ of the
desired reaction (e.g., a fission fragment for neutron-induced fission
reactions, or an emitted gamma-ray for capture reactions). The
experimental observable of interest is therefore the probability of this
coincidence
\begin{equation}
    P_{\chi}^{} (U) = \sum_{J,\Pi}
    F_{\delta\beta}^{}(U,J,\Pi) G_{\chi}^{}(U,J,\Pi),
    \label{eq:coinprob}
\end{equation}
where we have combined the formation probability of
Eq.~\eqref{eq:surrprob} with the branching ratio of Eq.~\eqref{eq:g}. As
mentioned above, the production of equilibrated compound-nuclear states
following the direct reaction is a non-trivial issue that requires
further attention. For the purpose of the present study we assume that
such states have been produced and that these states subsequently decay
according to the branching ratio $G_{\chi}^{}(U,J,\Pi)$.

In principle, width fluctuation corrections should be incorporated in
the above expression. However, if the formation of the \cn, represented
by $F_{\delta\beta}^{}(U,J,\Pi)$, is entirely independent of its
decay, i.e. has none or negligible contribution to the total decay
width, then the width fluctuation corrections can be applied to the
branching ratios alone. Surrogate reactions such as inelastic
scattering, are examples of this type of reaction. In this work, we will
neglect width fluctuations in the modeling of the surrogate decay
probabilities, but we will introduce them in the final step where we
extract the desired \ng\ cross section, see Eq.~\eqref{eq:surranalysis}.
\subsection{The angular momentum mismatch}
The relationship between Eqs.~\eqref{eq:desired} and \eqref{eq:coinprob}
constitutes the \hf\ formulation of the surrogate reaction method. While
the standard \hf\ formula expresses a cross section in terms of products
of formation cross sections and decay branching ratios, the experimental
observable in the surrogate approach is a coincidence decay probability
which is expressed in terms of formation probabilities and the decay
branching ratios. Although these two expressions contain the same
branching ratios, $G_{\chi}^{}(U,J,\Pi)$, they are weighted differently
because the $J\Pi$ distributions are different in the two reactions. In
fact, this mismatch can be quite significant. As we will show in
\sec{sec:surrogate}, the use of the surrogate technique to extract
low-energy \ng\ cross sections is particularly challenging since the
neutron will bring in very little angular momentum to the \cn\ compared
with that typically brought in by the $D(d,b)B^{*}$ direct reaction.  A
theoretical challenge in this case is therefore to determine the $J\Pi$
distribution $F_{\delta\beta}^{}(U,J,\Pi)$ so that the branching ratios
$G_{\chi}^{}(U,J,\Pi)$ can be extracted from the observed coincidence
decay probability, $P_{\chi}^{} (U)$, see Eq.~\eqref{eq:coinprob}.  If
this decomposition is well determined, the branching ratios can be
inserted in \eq{eq:desired} together with a formation cross section
$\sigma_{\alpha}^{}(\varepsilon_\alpha,J,\Pi)$ calculated with an
optical potential, to yield the desired $\sigma_{\alpha\chi}
(\varepsilon_\alpha)$ cross section. At this final stage we can also
introduce width fluctuation corrections, represented by the factor
$w_{\alpha\chi}$ in the following expression. This procedure for
analyzing a surrogate experiment can be outlined in diagram form
\renewcommand{\arraystretch}{1.5}
\beq
  \begin{array}{rlccc}
    \overbrace{P_{\chi}^{} (U)}^\mathrm{measured}
    &= \sum\limits_{J,\Pi}
    &\overbrace{F_{\delta\beta}^{}(U,J,\Pi)}^\mathrm{calculated} 
    &\overbrace{{G_{\chi}^{}(U,J,\Pi)}}^\mathrm{extracted} \\
    & & & \big\Downarrow \\
    \underbrace{\sigma_{\alpha\chi}(\varepsilon_\alpha)}_\mathrm{deduced} 
    &= \sum\limits_{J,\Pi}
    &\underbrace{\sigma_{\alpha}^{}(\varepsilon_\alpha,J,\Pi)}_\mathrm{calculated} 
    &{G_{\chi}^{}(U,J,\Pi)}
    &\underbrace{w_{\alpha\chi}(\varepsilon_\alpha,J,\Pi)}_\mathrm{calculated}.
  \end{array}
\label{eq:surranalysis}
\eeq
\renewcommand{\arraystretch}{1.0}
At this point we can remind ourselves that we have assumed that the
ejectile, $b$, is observed at a particular angle $\Omega_\beta$, or
alternatively that the experimental data is integrated over a limited
range of angles, and that we have neglected all angular dependence in
the formulas above. However, the formation probabilities
$F_{\delta\beta}^{}(U,J,\Pi)$ are most certainly angle-dependent, and
it is therefore desirable for surrogate experiments to be carried out
over a wide range of ejectile angles in the $D(d,b)B^{*}$
reaction. Obtaining the same $\sigma_{\alpha\chi}$ for the various
angles provides an important consistency check on the procedure.
%
%\subsubsection{The Weisskopf-Ewing limit}
%

An important simplification occurs when the branching ratios of
Eq.~\eqref{eq:g} do not depend on $J$ and $\Pi$. This removes most of
the model dependencies from the surrogate analysis since the
angular-momentum mismatch becomes irrelevant. This limit is known as the
Weisskopf-Ewing limit~\cite{wei40:57}. A number of conditions must be
satisfied in order for this limit to be
applicable~\cite{gad92:book,wei40:57,esc06:74}. For example, the energy
of the \cn\ must be sufficiently high so that all channels into which it
can decay are dominated by integrals over level densities. This
condition is not satisfied for low-energy neutron radiative capture and
the dramatic simplification of the surrogate analysis that occurs in the
Weisskopf-Ewing limit, see e.g. Ref.~\cite{bri79:72}, cannot be
utilized.  The breakdown of the validity of the Weisskopf-Ewing
approximation at low incoming neutron energies is illustrated in the
next section, see in particular \fig{fig:gbr}. Furthermore, the
spin-dependence of the level densities in the relevant channels has to
be of the simple form $\rho(U,J) \propto (2J+1)$. This condition is
satisfied if the spin $J$ is smaller than the spin-cutoff parameter
$\sigma_\mathrm{cut}$ in the relevant level-density formula. However, it
is known that the \we\ limit is still a useful approximation at somewhat
higher spins.
%
%**************************************************************************
\section{Neutron capture on Zr isotopes%
\label{sec:zr}}
In this section we present calculated \ng\ cross sections for a range of
Zr isotopes, obtained by applying \hf\ theory with decay models adjusted
to regional systematics. The purpose is threefold: Firstly, we want to
illustrate the quality of the developed systematics and present the
uncertainties connected to certain input parameters. Secondly, these
calculations provide a prediction for the unmeasured
$\nuc{95}{Zr}\ng\nuc{96}{Zr}$ cross section which is of importance for
stellar modeling efforts, see
e.g. Refs.~\cite{bus01:557,her05:43,lug03:586}. And finally, the
calculation presented here for \zrone \ng \zrtwo\ will serve as a
reference when discussing the feasibility of using surrogate experiments
to deduce \ng\ cross sections. We use our statistical-reaction decay
model to generate gamma-decay probabilities and we are able to perform
sensitivity tests by varying the most relevant input parameters.
\subsection{\label{subsec:decmod}Decay model and regional fit}
%-----------
%
% KK's piece of text included here:
% (modifications by CF)
%
Neutron capture cross sections modeled with the Hauser-Feshbach formula
are most sensitive to the photon transmission coefficients and nuclear
level densities. For the modeling of capture reactions with stable
targets, these two inputs can usually be constrained by experimental
data. In particular, the average total s-wave radiation width,
$\left<\Gamma_{\gamma}\right>_{0}$, may be used to determine the overall
normalization factor associated with the E1 photon strength function
while measured neutron s-wave resonance spacings, $D_{0}$, can be used
to fix level-density parameters at the neutron separation energy. For
cases where these quantities have not been measured one may attempt to
describe them by systematics.

A recent effort has been made to develop regional systematics (as
opposed to global prescriptions) for these two quantities for nuclei
spanning the range $34\le Z\le 46$ (selenium through
palladium)~\cite{hof06:tr}. The goal in developing these systematics was
to model several neutron- and charged-particle reactions on isotopes of
yttrium, zirconium, niobium, and molybdenum. Here we make use of the
findings of this work. Below we summarize the essential features of the
model. More details can be found in Ref.~\cite{hof06:tr}.

\begin{figure}[htb!]
\centering
\includegraphics*[width=80mm]{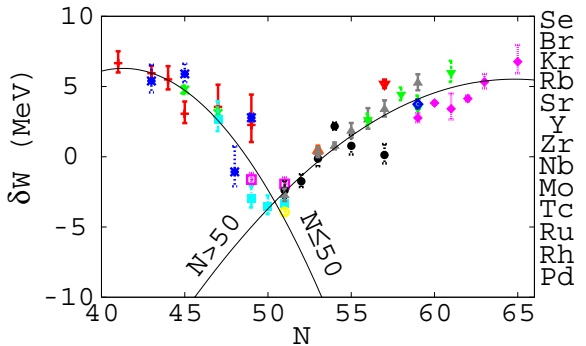}
\caption{\co\ Systematics for the shell correction used in the Fermi gas
description of the level density. See text for details.%
\label{fig:dw-sys}}
\end{figure}

The level density is described by a backshifted Fermi gas combined with
a constant temperature form at low excitation
energies~\cite{gil65:43,ilj92:543}. It is assumed that the parity
distribution of nuclear states is equal, i.e. for a given parity $\Pi$,
$\rho(U,J,\Pi) = \rho(U,J)/2$. The spin-dependence of the level-density
is described by
\beq
\label{eq:ldspin}
  f(U,J) = \frac{2J+1}{2\sigma^2} \exp{\left[ 
  \frac{-\left(J+\frac{1}{2}\right)^2}{2\sigma^2} \right]}, 
\eeq
where the spin-cutoff parameter is given by
\beq
\label{eq:spincutoff}
  \sigma^2 = 0.01496 \lambda A^{5/3} \sqrt{\frac{U}{a}}.
\eeq
In our analysis we fix $\lambda=1$. The level-density parameter $a$,
introduced above and also appearing in the Fermi gas formula, is
given an energy dependence
\beq
\begin{split}
\label{eq:ilj-apar}
  & a(U,Z,N)=\tilde{a}(Z,N) \\ 
  & \qquad \times \left[1+\delta W(Z,N) \frac{1-\exp\left[-\gamma
  (U-\Delta)\right]}{U-\Delta}\right]
\end{split}
\eeq
where $\tilde{a}(Z,N)$ is the asymptotic value for large $U$, $\delta
W(Z,N)$ is the shell correction as defined in~\cite{ilj92:543},
$(U-\Delta)$ is the backshifted energy, and $\gamma$ is a constant
factor.  Values for the backshifts are taken as the average
difference in binding energies of neighboring nuclei, as described
in~\cite{rau97:56}. Also taken from~\cite{rau97:56} are the values
$\tilde{a}(Z,N)=0.1337(Z+N)-0.06571(Z+N)^{2/3}$ and $\gamma=0.04884$
MeV$^{-1}$.  For nuclei with measured $D_{0}$, a shell correction
$\delta W(Z,N)$ is determined so that it will reproduce the measured
level spacing. For other nuclei, the shell correction is based on the
systematics shown in~\fig{fig:dw-sys}. These systematics were determined
by making two least-squares quadratic fits (one for either side of the
closed $N=50$ neutron shell) to the extracted shell corrections
\beq
\label{eq:dw-sys}
  \delta W(Z,N) = c_0 N^2 + c_1 N + c_2,% \qquad \mathrm{where}
\eeq
where
\begin{equation*}
\begin{array}{cccc}
  c_0= & c_1= & c_2= \\
  -0.114399 & 9.44901 & -188.821, & \quad \mathrm{for~} N \le 50 \\
  -0.0421006 & 5.45665 & -171.285, & \quad \mathrm{for~} N > 50, \\
\end{array}
\end{equation*}
valid for $34\le Z\le 46$. The error bars on the extracted shell
corrections reflect uncertainties in the measured $D_{0}$. 
The level-density parameters related
to the constant temperature form used at low excitation energies are
fixed by the known discrete spectrum and by requiring the two level
density formulae to match tangentially at an energy $U_{m}$. This
matching energy may be adjusted to provide the best possible fit to the
low-lying level structure without affecting the Fermi gas portion of the
level density. Such adjustments were made individually for each nucleus.

The systematics for the average total s-wave radiation width were
determined by assuming a simple dependence on the mass and s-wave
resonance spacing:
\begin{equation}
\label{eq:gg-sys}
  \left\langle \Gamma_{\gamma} \right\rangle_{0}^\mathrm{sys}=\left[aA +
  b\right]\left[c\log_{10}\left(D_{0}\right)+d\right].
\end{equation}
Making a least-squares fit to measured $\left\langle \Gamma_{\gamma}
\right\rangle_{0}$ (in meV) and $D_{0}$ (in keV) taken from the
compilations of~\cite{bel05:ripl} yields the coefficients $a=-5.928$,
$b=343.8$, $c=-42.43$, and $d=343.8$. 
These systematics were then used to generate a list of suggested values
of $\left\langle \Gamma_{\gamma} \right\rangle_{0}$ for nuclei in the
range $34\le Z\le 46$. The list includes the available experimental
values. Systematic values were chosen based on measured $D_{0}$ where
available, or on values of $D_{0}$ calculated from the regional level
density systematic shown in \fig{fig:dw-sys}. The line shape of the E1
photon strength function is described using a simplified version of the
extended generalized Lorentzian (EGLO) model~\cite{kop93:47}. The EGLO
model corresponds to a Lorentzian with an energy-dependent width. We are
using a simplified version in which the width is set to depend only on
the nuclear temperature at the energy of the decaying state and not on
the energy of the final state. The most important feature, however, is
that the magnitude of the strength function is normalized to reproduce
the average total s-wave radiation width.

Since we are interested in modeling reactions for a relatively large
region of isotopes we use the global nucleon-nucleus optical-model
potential by Koning and Delaroche~\cite{kon03:713} for the calculations
of the nucleon transmission coefficients. It has been
shown~\cite{kon03:713} that this parameterization gives a very
satisfactory fit to measured total cross section data and low-energy
observables, such as the s-wave strength function, that are relevant for
our application.
%
%-----------
\subsection{Neutron capture cross sections for Zr isotopes}
\begin{figure*}[phbt]
  \begin{minipage}{0.96\textwidth}
    \begin{minipage}[t]{0.49\textwidth}
      \centering
      \mbox{}\\
      \includegraphics*[width=\textwidth]{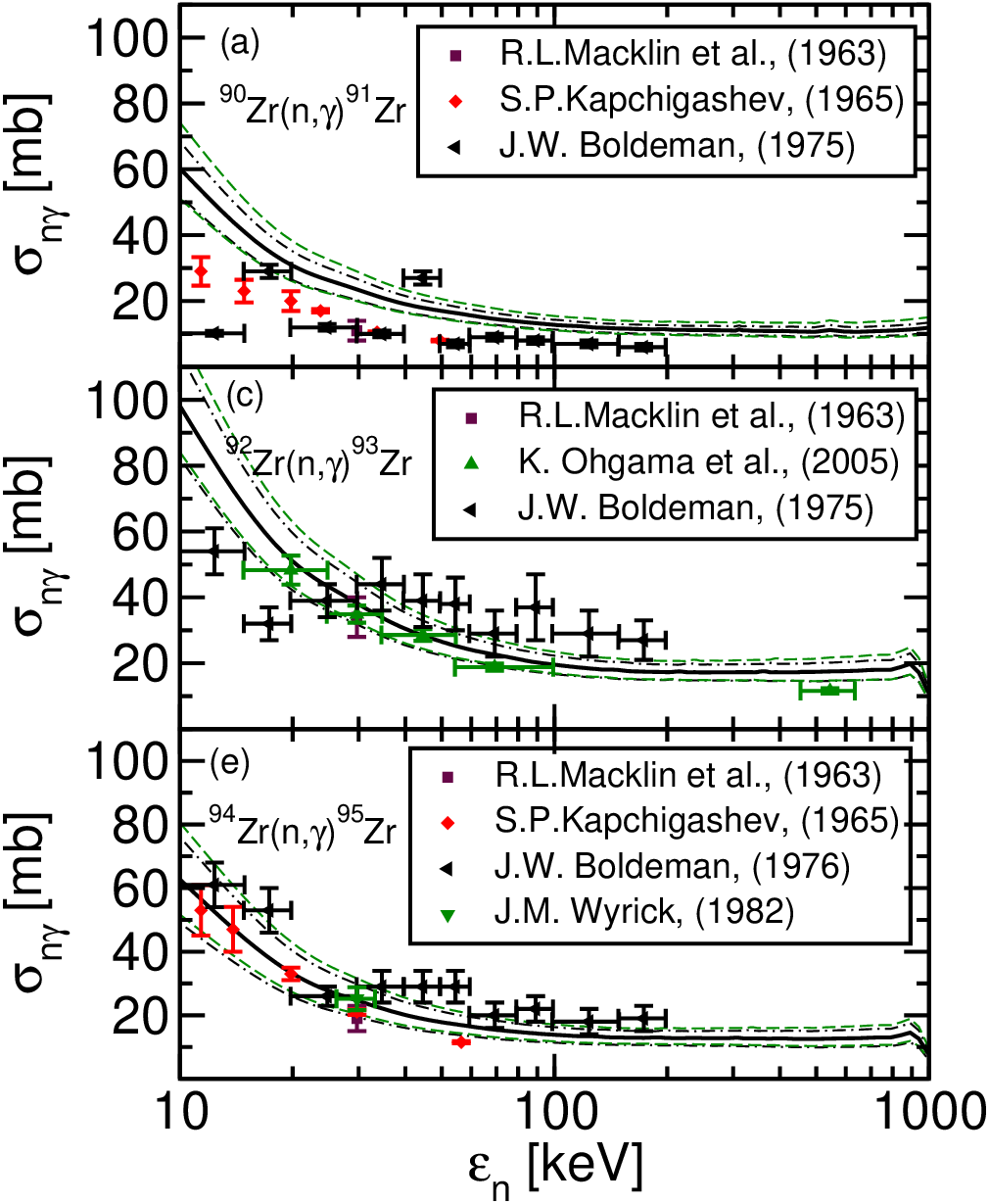}\\
      \mbox{}
    \end{minipage}
  \hfill
    \begin{minipage}[t]{0.49\textwidth}
      \centering
      \mbox{}\\
      \includegraphics*[width=\textwidth]{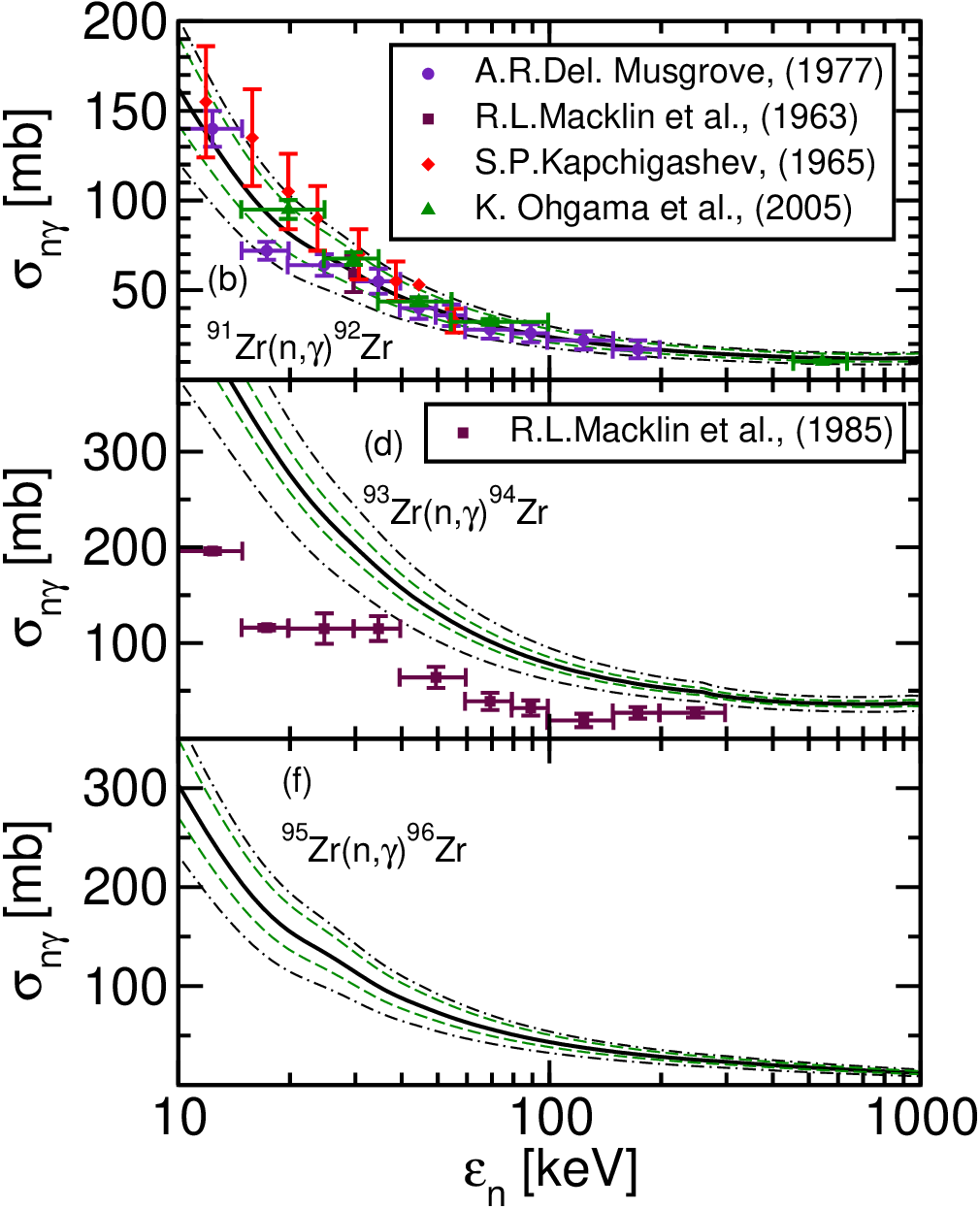}\\
      \mbox{}
    \end{minipage}
  \end{minipage}
  \caption{\co\ Calculated \ng\ cross sections compared to
  experimental data for six Zr isotopes: \zrzero, \zrone, \zrtwo, \zrthree,
  \zrfour, and \zrfive. Experimental data are from
  Refs.~\cite{mac63:8,kap65:19,bol75:246,
  mus77:30,ohg05:42,bol76:269,mac85:115,wyr82:tr,lyo59:114}. Dashed and
  dash-dotted lines indicate error bars associated with uncertainties in
  the s-wave resonance spacing $D_0$ and the s-wave average radiative width
  $\langle \Gamma_\gamma \rangle_0$, respectively. The error bars for
  the \zrthree\ and \zrfive\ \ng\ reactions are estimates based on
  typical experimental uncertainties for other isotopes in this region.
  \label{fig:ngcs}}
\end{figure*}

Our calculated \ng\ cross sections for the selected range of Zr isotopes
\nuc{90-95}{Zr} are shown in \fig{fig:ngcs}. The experimental data for
the quantities that constrained our decay model is summarized in
Table~\ref{tab:nghf}. As noted earlier, the two most important
ones are the average total s-wave radiation width,
$\left<\Gamma_{\gamma}\right>_{0}$, and the measured s-wave resonance
spacing, $D_{0}$. For the two radioactive isotopes, \zrthree\ and
\zrfive, for which no such experimental data exist, we have used values
obtained from the regional systematics described in the previous
section. The error bars given for these cases are estimates based on
typical experimental uncertainties for other isotopes in this
region. Error estimates for our calculated \ng\ cross sections are shown
as dashed and dash-dotted lines in \fig{fig:ngcs}. They were obtained by
repeating the calculations using the upper and lower limits of the
s-wave radiation width and level spacing, respectively. Our calculated
results are compared to available experimental data from the
EXFOR/CSISRS database~\cite{mac63:8,kap65:19,bol75:246,
mus77:30,ohg05:42, bol76:269, mac85:115, wyr82:tr,lyo59:114} and we find
a satisfactory agreement for the isotopes that are well studied and are
characterized by high level densities, which implies that the
statistical-reaction treatment is well justified. However, the resonance
spacing in \zrone\ is rather large and signatures of individual peaks
are observed in the $\zrzero \ng \zrone$ experimental data which
suggests that the statistical treatment might not be appropriate for
this case. In Table~\ref{tab:nghf} we also present the
Maxwellian-averaged \ng\ cross section (MACS) at $k_BT = 30$~keV. Our
results are compared to values recommended by Z.Y. Bao~\emph{et
al.}~\cite{bao00:76}, which are based on an evaluation of available
experimental and theoretical results.
\begin{table*}[phbt]
  \caption{Maxwellian-averaged \ng\ cross section at thermal energy $k_B
    T = 30$~keV calculated for different Zr targets. The results from
    this work are compared to the recommended values by Z.Y. Bao
    \emph{et al.}~\cite{bao00:76}. Also shown in the
    table are: $Q$ ($Q$-value for the reaction), $N_D$ (number of known
    discrete states in the \cn\ that form a complete spectrum), $U_m$
    (matching energy of constant temperature and Fermi-gas level-density
    regions), $D_0$ (level spacing for $s$-wave resonances at the
    neutron separation energy), and $\left\langle \Gamma_\gamma \right\rangle_0$
    ($s$-wave average radiative width). Experimental data are from the
    RIPL-2 database~\cite{bel05:ripl}, whereas unmeasured quantities are
    obtained from our regional systematics.
  \vspace*{1ex}%
  \label{tab:nghf}}
  \begin{ruledtabular}
    \begin{tabular}{lcccccc}
      & \zrzero\ng & \zrone\ng & \zrtwo\ng & \zrthree\ng & \zrfour\ng &
      \zrfive\ng \\
      \hline
      $Q$ [MeV]
      & 7.194 & 8.635 & 6.734 & 8.221 & 6.462 & 7.856 \\
      $N_D$ 
      & 41 & 6 & 3 & 9 & 1 & 8 \\
      $U_m$ [MeV] 
      & 3.167
      & 4.658 & 2.054 & 3.058 & 2.0 & 2.695 %\footnotemark[1] 
      \\
      $D_0$ [keV] 
      & $6.0 \pm 1.4$ & $0.55 \pm 0.10$ & $3.5 \pm 0.8$ & $0.160 \pm 0.015$ &
      $3.2 \pm 0.8$ & $0.26 \pm 0.04$\footnote[1]{There exists no direct
      experimental information for this observable. The given value is based on
      the systematics from our regional fit.}\\
      $\left\langle \Gamma_\gamma \right\rangle_0$ [meV] 
      & $130 \pm 20$ & $140 \pm 40$ & $135 \pm 25$ & $164 \pm
      40$\footnotemark[1] & $85 \pm 20$ & $144 \pm 40$\footnotemark[1]\\
      \hline
      \multicolumn{7}{c}{Maxwellian-averaged \ng\ cross section (in mb)
        at thermal energy $k_B T = 30$~keV}\\
      \hline
      This work
      \footnote[2]{The error bar given here is $\sqrt{\Delta_D^2 +
      \Delta_\Gamma^2}$, where $\Delta_{D,\Gamma}$ are the errors due to the
      uncertainties in $D_0$ and $\langle \Gamma_\gamma \rangle_0$,
      respectively.}
      & $26 \pm 7$ & $62 \pm 19$ & $41 \pm 11$ & $198 \pm
      44$ & $27 \pm 9$ & $ 117 \pm 34$\\
      Bao \emph{et al.}~\cite{bao00:76}
      & $21 \pm 2$ & $60 \pm 8$ & $33 \pm 4$ & $95 \pm 10$ & $26 \pm 1$
      & $79 \pm 12$\footnote[3]{This recommendation is based purely on
      theoretical estimates.}
    \end{tabular}
  \end{ruledtabular}
\end{table*}
\begin{table}[hbt]
  \caption{A summary of the decay models for \zrone\ng\zrtwo\  that are
  used in this work to investigate the sensitivity to certain key
  quantities.
  \vspace*{1ex}% 
  \label{tab:decmod}}
  \begin{ruledtabular}
    \begin{tabular}{lcccc}
      & $\left\langle \Gamma_\gamma \right\rangle_0$
      & $D_0$ & $\delta W$ & MACS at \\
      & [meV]
      & [keV] & [MeV] & 30~keV [mb] \\
      \hline
      Reference
      & 140 & 0.55 & -1.743 & ~62 \\
      Decay model 1
      & 280 & 0.55 & -1.743 & 115 \\
      Decay model 2
      & 140 & 0.45 & -1.237 & ~74
    \end{tabular}
  \end{ruledtabular}
\end{table}
%
%
%**************************************************************************
\section{Theoretical analysis of decay data from a surrogate experiment%
\label{sec:surrogate}}
In this section we explore different approaches for utilizing decay
probabilities measured in a surrogate experiment to extract low-energy
neutron capture cross sections. We perform our simulations for
the \zrone \ng \zrtwo\ reaction. Let us first take a closer look at some
of the details of the statistical-reaction calculation introduced in the
previous section. The decay model for \zrtwo\ described in
\sec{subsec:decmod} will act as our reference model for these studies,
and the corresponding \ng\ cross section serves as the reference
(or ``true'') cross section that we seek to extract from our
simulated surrogate experiments.
%-----------
\subsection{The \zrone \ng \zrtwo\ reaction%
\label{sec:sensitivity}}
\begin{figure}[phbt]
  \includegraphics*[width=80mm]{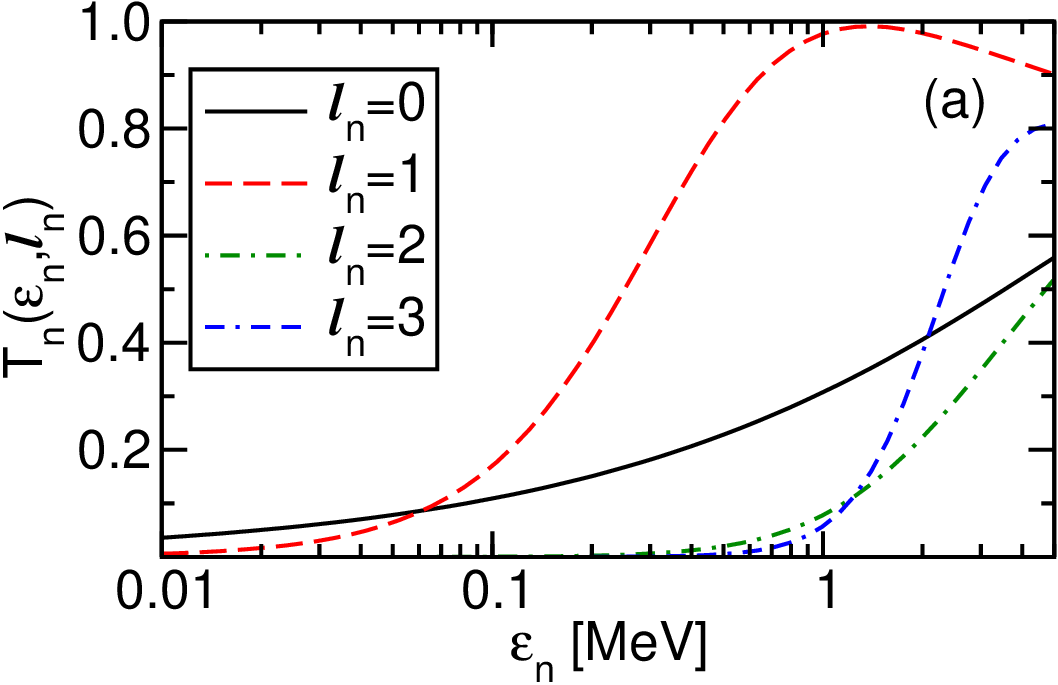}\\[2ex]
  \includegraphics*[width=80mm]{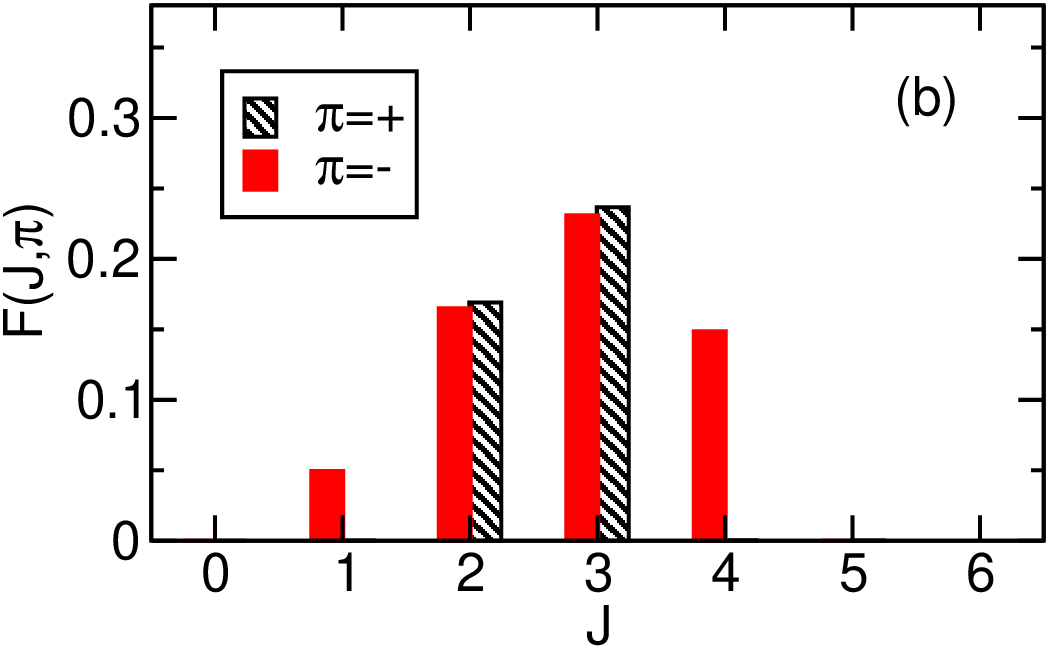}\\  
  \caption{\co\ (a) Energy- and $l$-dependence of neutron transmission
  coefficients on a \zrone\ target. (b)
  $J\Pi$ population of the \zrtwo\ \cn\ following neutron absorption on
  $\zrone(5/2^+)$ at $\varepsilon_n = 30$~keV.%
  \label{fig:n_jpi_pop}}
\end{figure}

Neutron transmission coefficients calculated with the global
optical-model potential by Koning and Delaroche~\cite{kon03:713} are
shown in \fig{fig:n_jpi_pop}(a) for several partial waves on a \zrone\
target. For $l>0,$ these are the appropriately weighted averages of the
coefficients with $j=l \pm \frac{1}{2}$. The rapid increase of the
p-wave transmission coefficient with increasing neutron energy is
characteristic of all isotopes in this mass region, and corresponds to
the well known $3p$ giant single-particle resonance. An interesting
consequence of this feature is that low-energy neutron absorption (above
approximately 10~keV) populates predominantly the $J\Pi$-states of the
\cn\ that can be reached by either s-wave or p-wave capture. The effect
is clearly illustrated in \fig{fig:n_jpi_pop}(b) where the $J\Pi$
population of \zrtwo, following the absorption of 30~keV neutrons on
$\zrone(5/2^+)$, is shown to be dominated by $1^-,2^-,3^-,4^-$ and
$2^+,3^+$ states.

It is important to realize that there is generally a very small number
of open decay channels following the absorption of low-energy neutrons
on spherical, or near-spherical, targets. Other particle-decay
thresholds usually appear above the neutron separation energy, and
inelastic channels are closed when the first excited states of the
target are above the incoming neutron energy. This leaves only two
possible decay paths: compound-elastic scattering leading back to the
target ground state, and gamma deexcitation of the \cn. We now focus on
the branching ratio defined in \eq{eq:g}. In this special case, the sum
over open channels ($\chi'$) appearing in the denominator contains only
two terms. The important quantities are therefore the neutron
transmission coefficients discussed above and the product of the gamma
transmission coefficients with the level density of the \cn. The latter
quantity must be integrated over all states that can be reached by a
primary gamma transition. For low-energy neutrons, a single gamma
deexcitation is usually enough to bring the \cn\ below the
neutron-decay threshold, which will inevitably lead to a continued gamma
cascade down towards the ground state. As noted previously, the
\textsc{Stapre} code correctly computes the gamma cascade even when
neutron emission is possible within the cascade.

\begin{figure}[phbt]
  \includegraphics*[width=80mm]{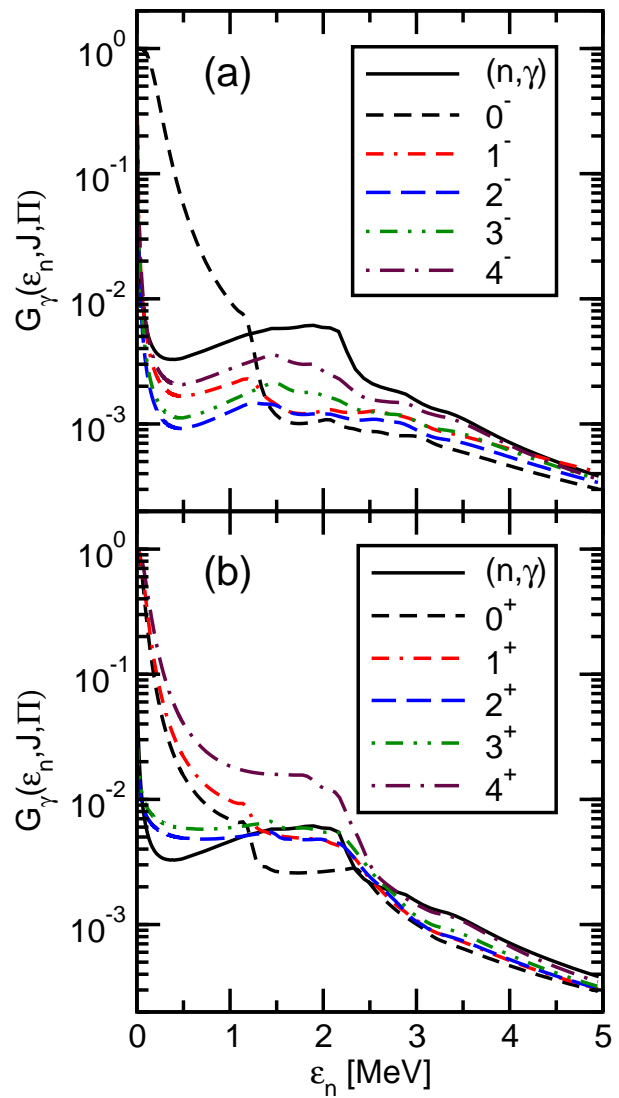}\\
  \caption{\co\ Gamma branching ratios as a function of spin, parity and
  excitation energy of the decaying state in \zrtwo. The energy is given
  as the equivalent neutron energy $\varepsilon_n = U - S_n$. The total
  gamma-decay probability following neutron absorption, as described in
  the text, is shown as a solid line in both panels. Width-fluctuation
  corrections were turned off when producing these results.%
  \label{fig:gbr}}
\end{figure}

Fig.~\ref{fig:gbr} shows the gamma branching ratios for \zrtwo\ as a
function of $J\Pi$ and neutron energy. For comparison we also plot the
total gamma decay probability which we define as the \zrtwo\ production
cross section divided by the neutron absorption cross section. These
results are obtained with width-fluctuation corrections turned off. The
results shown in this plot demonstrate a feature of nuclei that is
particularly striking near closed shells, namely a dramatic dependence
of the compound-nucleus gamma decay branching ratios on $J\Pi$ at low
energies above threshold. For some $J\Pi$ values the probability for
gamma decay is close to one, whereas for others it is on the order of
$10^{-3}$. Clearly we are very far from the \we\ limit and certain
approximations that have previously been used in the analysis of
surrogate experiments cannot be used here. The reason for the dramatic
dependence on $J\Pi$ is the behavior of the neutron transmission
coefficients, shown in \fig{fig:n_jpi_pop}(a), in combination with the
small number of discrete states at low excitation energies in the
initial nucleus \zrone. Those compound-nuclear states in \zrtwo\ that
can reach an energetically allowed state in \zrone\ by the emission of
an s- or p-wave neutron have a very small gamma-decay probability. For
the other states, the branching ratios are determined by the competition
between $l \ge 2$ neutron transmission coefficients and gamma
transmission coefficients. It turns out that the gamma-decay channel
usually dominates although the gamma transmission coefficients are still
very small. It is also important to note the convergence of the curves
at higher energies $E_n \gtrsim 3$~MeV. This demonstrates the onset of
the Weisskopf-Ewing limit with the resulting approximate independence of
$J\Pi$ for the branching ratios. In the \sec{sec:analysis} we return to
this issue and discuss how we can utilize the observed behavior for our
purposes.

Before we discuss the sensitivity to properties of the nuclear decay
model, we introduce the notation
$\sigma^\rmref_{n\gamma}(\varepsilon_n)$,
$\sigma^\rmref_\mathrm{abs}(\varepsilon_n)$, and
$G^\rmref_\gamma(U,J,\Pi)$. These quantities are the results of our
reference calculation and they denote, respectively, the radiative
neutron-capture cross section, the neutron absorption cross section, and
the gamma branching ratios of Eq.~\eqref{eq:desired}.

\begin{figure}[phbt]
  \includegraphics*[width=80mm]
		   {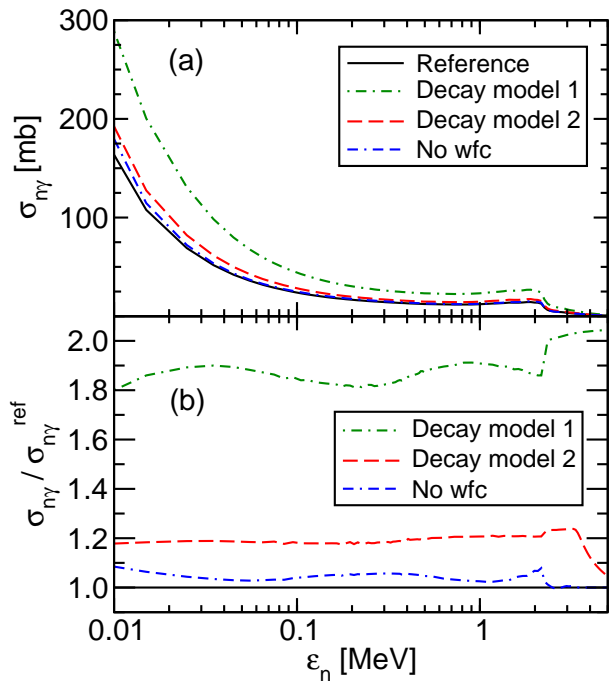}\\
  \caption{\co\ Sensitivity of the \zrone\ng\zrtwo\ cross section to
  variations of key parameters in the decay model. (a) Cross sections
  calculated using the three decay models of Table~\ref{tab:decmod} plus
  one calculation using the Reference Decay model but turning off
  width-fluctuation corrections. (b) The ratio to the reference cross
  section.%
  \label{fig:ngcs_sens}}
\end{figure}

By introducing variations from the reference decay model we will be able
to study the effects of our lack of knowledge of the true decay model,
which is one of the two major uncertainties associated with the
surrogate method.  As previously mentioned, the modeling of neutron
capture reactions is most sensitive to the nuclear level density at, and
somewhat below, the neutron separation energy, as well as to the E1
photon strength function at small gamma energies ($\sim 2-3$~MeV). In
order to quantitatively investigate this sensitivity we introduce two
decay models, ``Decay model 1'' and ``Decay model 2'', in addition to
the ``Reference Decay model'' corresponding to the parameter set from the
regional systematics. The most important parameters for the three models
are summarized in Table~\ref{tab:decmod}, together with the resulting
MACS at 30~keV. The salient feature of Decay model 1 is a factor of two
increase in the magnitude of the E1 photon strength function. Such
uncertainty in the size of the gamma-transition strength is not
unrealistic when moving to unstable isotopes for which no experimental
data exists. The \ng\ cross section is almost directly proportional to
the magnitude of the E1 photon strength function, as is clearly seen
in Fig.~\ref{fig:ngcs_sens} where the ratio of the modeled capture cross
section to the reference calculation is shown. The logarithmic energy
axis was introduced in order to cover a large energy range while still
focusing on the relevant small energies. For Decay model 2 we decreased
the level spacing $D_0$ for s-wave resonances by modifying the shell
correction $\delta W$ in the formula for the level-density parameter
$a$, Eq.~\eqref{eq:ilj-apar}. We choose this particular way of modifying
the level density since $\delta W$ is the free parameter in our regional
systematic. The $\lesssim 20$\% decrease of the level spacing at the
separation energy implies a $\gtrsim 20$\% increase of the level
density, and again we observe a proportional increase in the capture
cross section. Finally, we present the results obtained when turning off
the width-fluctuation corrections. As expected, turning off
width-fluctuation corrections results in an increase in the capture
cross section. From Fig.~\ref{fig:ngcs_sens} we infer that it is
approximately a 10\% effect at small energies and completely negligible
at higher energies ($\gtrsim 2$~MeV).
%
%
%-----------
\subsection{Simulating the direct reaction of a surrogate experiment
\label{sec:simsurr}}
The first step of the reaction that takes place in a surrogate
experiment is a direct reaction, such as a transfer or inelastic
scattering reaction, that produces the relevant intermediate nucleus in
a highly excited state. For the purpose of measuring the decay
probabilities that are pertinent for the desired reaction, it is
important that the intermediate nucleus first equilibrates into a
compound-nuclear state. The relevant net result is therefore the
distribution of $J\Pi$ states in the \cn\ that is populated following
the direct reaction. In principle, one would like to be able to describe
this direct-reaction process, leading to an equilibrated \cn, using an
appropriate direct-reaction model. However, this is a non-trivial task
since it requires a description of transfer and inelastic scattering
reactions leading to unbound states, as well as an understanding of the
damping of those states into equilibrated compound-nuclear
states. Moreover, a variety of projectile-target combinations with a
range of possible incident energies may be considered for producing the
\cn\ of interest. Different reaction mechanisms, regions of the nuclear
chart, and projectile energies yield different compound-nuclear $J\Pi$
distributions and also provide different challenges for a proper
theoretical description. DWBA calculations relevant to the present work
for inelastic alpha scattering to highly excited states in spherical
nuclei are currently under way and will be reported elsewhere.

\begin{figure}[phbt]
  \includegraphics*[width=80mm]{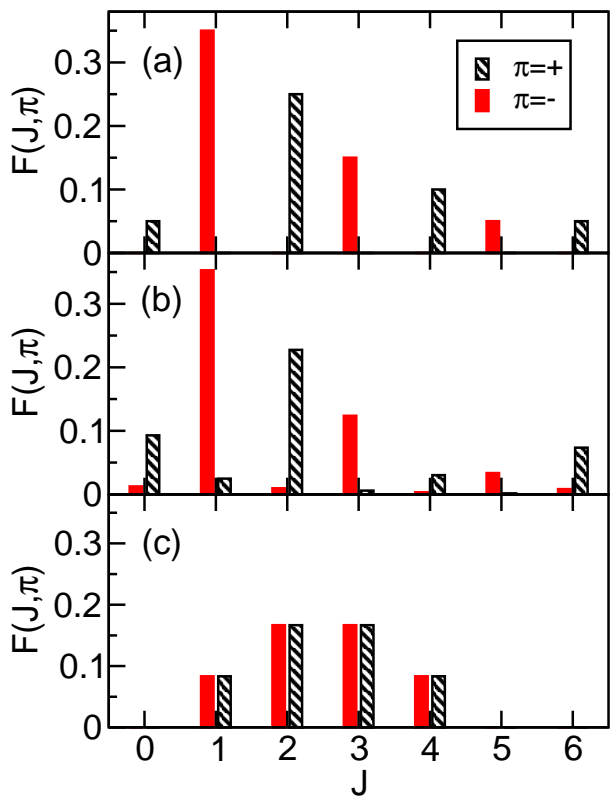}
  \caption{\co\ Schematic surrogate populations used in this study: (a)
  $F^{A,\rmref}(J,\Pi)$; (b) $F^{A+\delta}(J,\Pi)$; (c)
  $F^B(J,\Pi)$. See text for further details.%
  \label{fig:pops}}
\end{figure}

For the present purpose of simulating surrogate experiments we use
simple schematic distributions. Furthermore, we assume that the
distributions are independent of the \cn\ excitation energy in the
energy range just above the neutron separation energy that we are
interested in. Since inelastic scattering reactions can be used to
populate most compound nuclei that are relevant for studies of
s-process branch points, we employ distributions that exhibit the
asymmetry between natural- and unnatural-parity states that is
characteristic for those reactions with even-even targets. In the
absence of a spin-dependent interaction, a distorted-wave Born
approximation description of such a reaction yields only $J^\Pi = 0^+,
1^-, 2^+, 3^-, \ldots$ states. Distribution A, shown in
\fig{fig:pops}(a), contains only such natural-parity states and we will
use this distribution as the reference (``true'') population of the \cn\
\zrtwo$^*$ in a simulation of our benchmark surrogate reaction
$\zrtwo(\alpha,\alpha')\zrtwo^*$. For this reason we denote this
distribution $F^{A,\rmref}(J,\Pi)$. Furthermore, the gamma-decay
probability that is obtained when combining this $J\Pi$ distribution
with our reference gamma branching ratio, $G^\rmref_\gamma(U,J,\Pi)$, in
Eq.~\eqref{eq:coinprob} will be denoted $P^\rmref_\gamma(U)$. The decay
probability $P^\rmref_\gamma(U)$ thus corresponds to what would be
measured in our simulated surrogate experiment, indicated as a solid
curve in \fig{fig:sim3}(a).

In addition to $F^{A,\rmref}(J,\Pi)$ we use two more schematic
distributions: $F^{A+\delta}(J,\Pi)$ and $F^B(J,\Pi)$. They are shown in
\fig{fig:pops}(b,c), respectively. The first one is almost identical to
$F^{A,\rmref}(J,\Pi)$, but contains a small random noise to simulate a
minor error in the predicted distribution.
Finally, Distribution B represents a significantly different population
without any asymmetry between different parity states. The average
angular momentum deposited in the compound nucleus, $\langle J \rangle$,
is approximately 2.0 for all distributions. Therefore, these simulations
correspond to a study of the sensitivity to deviations in the assumed
$J\Pi$ distribution while keeping the average transferred angular
momentum approximately fixed.
%
%-----------
\subsection{Three alternative analysis approaches%
\label{sec:analysis}}
In the following we will assume that Distribution A,
$F^{A,\mathrm{ref}}(J,\Pi)$, is a reasonable representation of the
$J\Pi$ distribution of a compound nucleus created in a surrogate
reaction (such as inelastic scattering off an even-even near-spherical
nucleus). Also, we assume that the Reference Decay model with the
parameters from Section~\ref{sec:zr} represents a realistic description
of the ``true'' decay of the \cn, populated either through the
neutron-induced or the surrogate reaction. The function
$P_\gamma^\mathrm{ref}(U)$, calculated from \eq{eq:coinprob} using the
(energy-independent) formation probability $F(J,\Pi)$ from Distribution
A and the branching ratios $G_{\chi}^{}(U,J,\Pi)$ from the Reference
Decay model, is then the quantity that is observed in our simulated
Surrogate experiment.  We now investigate various possibilities of
extracting the desired ``true'' cross section
$\sigma_{n\gamma}^\mathrm{ref}(\varepsilon_n)$, represented by the
``Reference'' cross section in \fig{fig:ngcs_sens}. In order to test the
different procedures, we introduce uncertainties in the modeling by
using Decay model 1 or 2 (see Table~\ref{tab:decmod}).  This will
illustrate the effect of having insufficient knowledge of the ``true''
decay of the compound nucleus. Furthermore, for the approaches that
require the theoretical prediction of the $J\Pi$ population of the \cn\
created in the surrogate reaction we make use of the schematic
distributions introduced in Sec.~\ref{sec:simsurr}.  This will
illustrate the effect of having insufficient knowledge of the $J\Pi$
distribution of the decaying nucleus.

We discuss three different approaches that utilize decay data from a
surrogate experiment to extract the desired low-energy \ng\ cross
section. The three approaches are labeled as follows:
\begin{enumerate}
  \item{\we\ approximation.}
  \item{Full modeling of the $J\Pi$ population following the surrogate
  reaction.}
  \item{Normalization of the decay model in the \we\ region.}
\end{enumerate}
Below we discuss each one of these approaches in some detail and draw
some conclusions regarding their applicability.
\subsubsection{\we\ approximation}
The results and discussion of the preceding sections imply that the
convenient \we\ approximation cannot be utilized in the surrogate
approach when trying to extract low-energy \ng\ cross sections for
spherical and near-spherical targets. It was shown that the gamma-decay
branching ratios depend very sensitively on the particular $J\Pi$
population of the intermediate nucleus due to the small number of open
decay channels.

\begin{figure}[phbt]
  \includegraphics*[width=80mm]
    {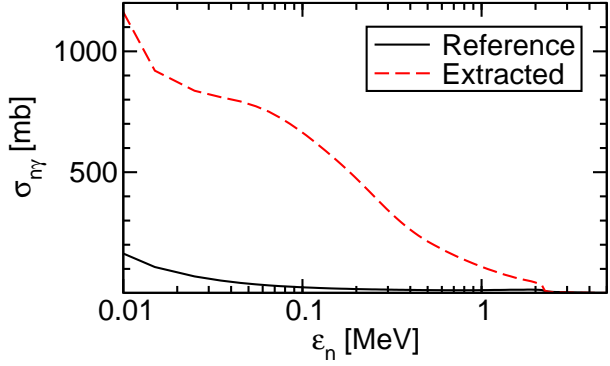}
  \caption{\co\ Extraction of \ng\ cross section from a surrogate
  experiment simulation using the \we\ approximation; see simulation
  ``WE'' of Table~\ref{tab:sim} and discussion in text.%
  \label{fig:sim1}}
\end{figure}

Fig.~\ref{fig:sim1} illustrates the inadequacy of the \we\ approximation
for this purpose. In this simulation the extracted \ng\ cross section is
obtained by simply multiplying the reference absorption cross section
[$\sigma^\rmref_\mathrm{abs}(\varepsilon_n)$] by the reference surrogate
decay data, [$P^\rmref_\gamma(U)$]. This procedure corresponds to the
simulation labeled ``WE'' in Table~\ref{tab:sim}. The extracted cross
section is compared to the reference cross section. We find
disagreement at the level of one order of magnitude, and also that the
shape of the extracted cross section is wrong.
\begin{table}[hbt]
  \caption{Combinations of decay models (see Table~\ref{tab:decmod}) and
  $J\Pi$ distributions (see \fig{fig:pops}) used in the simulations. 
  \vspace*{1ex}% 
  \label{tab:sim}}
  \begin{ruledtabular}
    \begin{tabular}{ccc}
      Simulation & Decay model & $J\Pi$ distribution \\
      \hline
      Reference & Reference & A,ref \\
      WE & Reference & --- \\
      1 & 2 & A+$\delta$ \\
      2 & 2 & B \\
      3 & 1 & A+$\delta$ \\
      4 & 1 & B
    \end{tabular}
  \end{ruledtabular}
\end{table}

We briefly mention here the surrogate ratio method, which is based on
the \we\ approximation. This method has been used to extract $(n,f)$
cross sections in the actinide region~\cite{bur06:73,ple05:71}. The goal
of the ratio method is to determine the ratio of two cross sections of
two ``similar'' compound-nuclear reactions.  An independent
determination of one of these cross sections then allows one to infer
the other. The desired ratio of the two cross sections is obtained
indirectly from surrogate measurements of the ratio of decay
probabilities under the assumption that the \we\ limit is valid. It was
demonstrated in Ref.~\cite{esc06:74} that the ratio approach may
actually reduce the error that is usually associated with neglecting the
$J\Pi$ dependence of the branching ratios. However, as we have just
shown in Fig.~\ref{fig:sim1}, the error encountered when applying the
simple \we\ approximation to low-energy \ng\ reactions is about one
order of magnitude larger than in the $(n,f)$ case. In the latter case,
the error rarely exceeds a factor of two even at small neutron energies
(see in particular Fig.~10 of Ref.~\cite{esc06:74}). Furthermore, for
spherical or near-spherical targets, the $J\Pi$ dependence of the
gamma-decay branching ratios at low energies is very sensitive to the
particular level structure of the target, as shown in
\sec{sec:zr}. Thus, it is hard to reduce the effects of this sensitivity
simply by measuring the ratio of decay probabilities for two neighboring
isotopes.
\subsubsection{Full modeling of the $J\Pi$ population following the
  surrogate reaction%
\label{sec:sim2}}
For applications in the actinide region there have been attempts to take
the effects of the $J\Pi$ population mismatch into account when
estimating $(n,f)$ cross sections from surrogate experiment
data~\cite{you03:67,you03:68}. In that work, the population of the \cn\
following a $(t,p)$ reaction was calculated in a distorted-wave Born
Approximation approach. Furthermore, efforts are currently underway to
develop a more advanced direct-reaction framework in which the surrogate
reactions to unbound states, and their subsequent equilibration into
compound-nuclear states, can be studied in greater detail. However, the
application of the surrogate method to low-energy \ng\ reactions is very
challenging, in particular when near-spherical targets are involved. We
will show that a small uncertainty in the predicted $J\Pi$ population
can lead to a very large error in the extracted cross section.

A surrogate analysis inspired by the diagram depicted in
Eq.~\eqref{eq:surranalysis} is performed by first introducing a
modeled gamma-decay probability, $P_{\gamma}^\mathrm{model}(U)$, with
both the $J\Pi$ distribution and the gamma-decay branching ratios
obtained from initial modeling efforts
\beq
\begin{split}
    P_{\gamma}^\mathrm{model} (U) &= \sum_{J,\Pi}
    F^\mathrm{model}(U,J,\Pi)
    G_{\gamma}^\mathrm{model}(U,J,\Pi) \\
    &\equiv \sum_{J,\Pi} P_{\gamma}^\mathrm{model}(U,J,\Pi). 
    \label{eq:Pmodel}
\end{split}
\eeq
This calculated quantity is then compared to the measured gamma-decay
probability. A fit to the experimental data is
achieved by introducing a fitting function $\eta(U,J,\Pi)$ that
relates the modeled decay probability with the measured one
\begin{equation}
    P_{\gamma}^\mathrm{exp} (U) = \sum_{J,\Pi}
    \eta(U,J,\Pi)
    P_{\gamma}^\mathrm{model}(U,J,\Pi). 
    \label{eq:fitetaUJP}
\end{equation}
For simplicity we assume that the functions $\eta(U,J,\Pi)$ are
independent of $J$ and $\Pi$ so that the fit corresponds to
an energy-dependent normalization. In this case it is simple to evaluate
the correction factor as
\begin{equation}
    \eta(U) = 
    \frac{P_{\gamma}^\mathrm{exp} (U)}
    {P_{\gamma}^\mathrm{model} (U)}, 
    \label{eq:fitetaU}
\end{equation}
and subsequently use it to extract the desired cross section
\beq
\begin{split}
    \sigma^\mathrm{extract}_{n\gamma}(\varepsilon_n)
    = \eta(U) & \sum_{J,\Pi}
    \sigma^\mathrm{model}_\mathrm{abs}(\varepsilon_n,J,\Pi) \\
    &\quad \times G^\mathrm{model}_\gamma(U,J,\Pi)
    w_{n\gamma}(\varepsilon_n,J,\Pi).
    \label{eq:sigsim2}
\end{split}
\eeq
In the final step we have also introduced the theoretical
width-fluctuation corrections, $w_{n\gamma}(\varepsilon_n,J,\Pi)$. The
numerical calculations discussed below were calculated using the
statistical-model code \textsc{Stapre} to calculate the various factors in
the above equation, including the width-fluctuation corrections,
$w_{n\gamma}(\varepsilon_n,J,\Pi)$. 

\begin{figure*}[phbt]
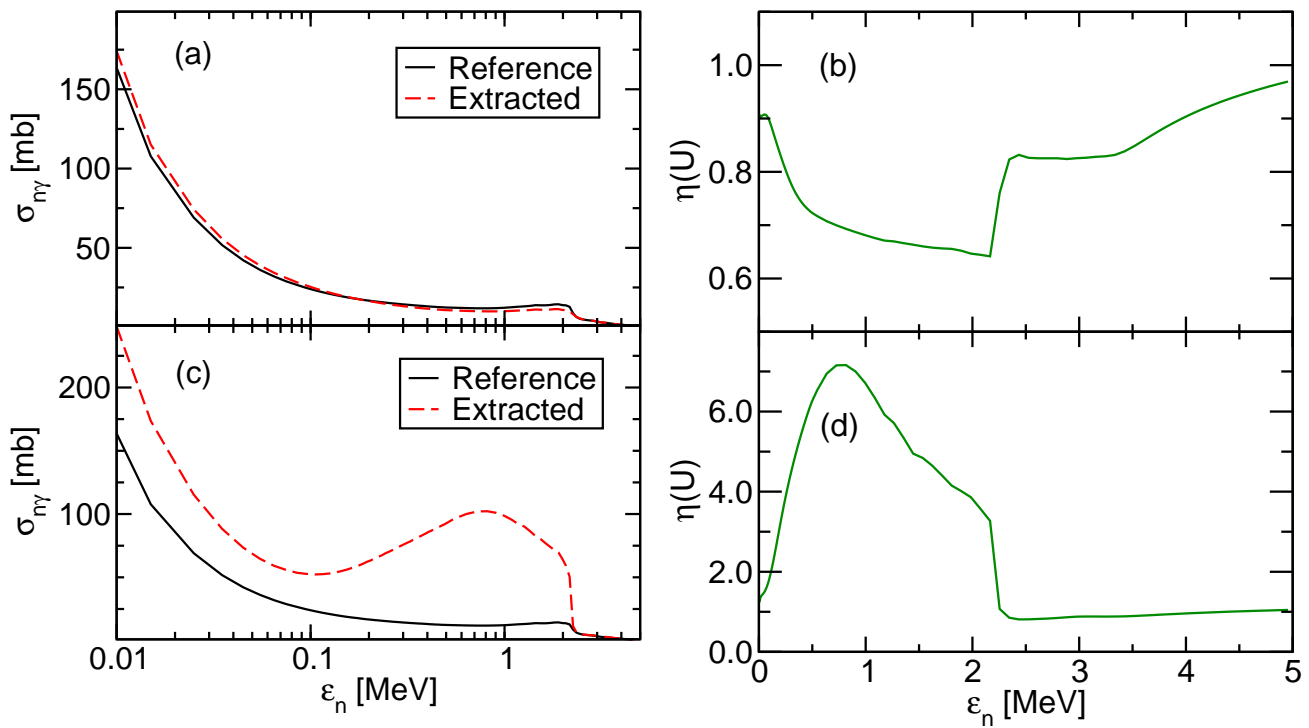

  \begin{minipage}{0.96\textwidth}
    \begin{minipage}[t]{0.49\textwidth}
      \centering
      \mbox{}\\
      \includegraphics*[width=\textwidth]
		       {fig8a_Zr91-GS-ng_surranalysis_approach2xs.eps}\\
      \mbox{}
    \end{minipage}
  \hfill
    \begin{minipage}[t]{0.49\textwidth}
      \centering
      \mbox{}\\
      \includegraphics*[width=\textwidth]
		       {fig8b_Zr91-GS-ng_surranalysis_approach2eta.eps}\\
      \mbox{}
    \end{minipage}
  \end{minipage}
  \caption{\co\ Extraction of \ng\ cross section from two different
  surrogate experiment  simulations using a full modeling of the $J\Pi$
  population: (a)  Simulation 1, and (c) Simulation 2, see
  Table~\ref{tab:sim}. Panels (b) and (d) show the respective
  normalization functions $\eta(U)$. Note that $U = \varepsilon_n +
  S_n$.%
  \label{fig:sim2}}
\end{figure*}

We will now test this procedure by performing several simulations. We
use the simulated surrogate decay data, $P_{\gamma}^\mathrm{ref}(U)$, as
corresponding to the experimental data in the numerator of
\eq{eq:fitetaU}. As for the calculated quantity appearing in the
denominator, we introduce some variations in the modeled decay probabilities
and the predicted $J\Pi$ population. The question is to what extent the
analysis procedure will be able to correct for these variations.  In the
first simulation we use Decay model 2 to calculate
$G_{\gamma}^\mathrm{model}(U,J,\Pi)$. We know that this decay model
overestimates the cross section by $\sim$20 \% (see
Table~\ref{tab:decmod}).  Furthermore, we assume that we have a very
good, but not perfect, estimate of the $J\Pi$ population of the
intermediate nucleus. The distribution $F^\mathrm{model}(U,J,\Pi)$ only
differs from the reference (``true'') population $F^{A,\rmref}(J,\Pi)$
by a small random noise. For this purpose we use the distribution
$F^{A+\delta}(J,\Pi)$ shown in Fig.~\ref{fig:pops}(b). The combination
of Decay model 2 and the $J\Pi$ distribution $F^{A+\delta}(J,\Pi)$ is
denoted Simulation 1 in Table~\ref{tab:sim}. The procedure outlined
above results in the normalization function $\eta(U)$, shown in
\fig{fig:sim2}(b) plotted as a function of the equivalent neutron energy
$\varepsilon_n$. Using this normalization function in \eq{eq:sigsim2}
yields the extracted \ng\ cross section shown as a dashed line in
\fig{fig:sim2}(a). For this case the procedure works relatively well and
the MACS at 30~keV for the extracted cross section is 67~mb. This number
should be compared with the 74~mb that is the original prediction of
Decay model 2, and the 62 mb that is the reference result that we were
aiming for.

In the second simulation we use a model for the $J\Pi$ population that
is a poor representation of the ``true'' $J\Pi$ distribution.  In this
case, we find that the approach outlined above results in a very poor
correction of the decay model, and consequently of the extracted cross
section. For this simulation we use the combination of Decay model 2 and
the $J\Pi$ distribution $F^{B}(J,\Pi)$, which is denoted Simulation 2 in
Table~\ref{tab:sim}. The extracted \ng\ cross section is shown as a
dashed line in \fig{fig:sim2}(c), and the normalization function
$\eta(U)$ is shown in \fig{fig:sim2}(d). The extracted cross section is
very different from the desired result below about 2.2~MeV.

The above results imply that one may be able to use surrogate data to
correct a theoretical decay model provided one has sufficiently accurate
information on the $J\Pi$ population of the \cn\ that decays in the
surrogate experiment.  Obtaining a reliable prediction of the relevant
$J\Pi$ population is challenging, since it requires accurate
direct-reaction calculations involving the nuclear
continuum. Furthermore, the equilibration process that follows the
production of a highly excited, intermediate nucleus in a surrogate
reaction is not sufficiently well understood; possible decay mechanisms
other than damping into the compound nucleus need to be accounted for if
they are present.  Nevertheless, it may be possible to obtain some
experimental signatures of the $J\Pi$ population of the decaying
nucleus. While the high sensitivity of the gamma-decay branching ratios
to the $J\Pi$ population makes the extraction of the desired cross
section from a surrogate experiment very difficult, it should also
result in certain experimental observables, such as the relative
intensities of discrete gamma transitions, being useful to constrain
calculated $J\Pi$ distributions.  This remains to be investigated in
more detail.
\subsubsection{Normalization of the decay model in the \we\ region%
\label{sec:sim3}}
In the third approach we utilize the fact that surrogate experiments
can, in principle, provide decay data for a very wide energy range. Thus
it is possible to collect data from an energy region in which the \we\
limit is approximately correct. For \zrtwo, we find from \fig{fig:gbr},
that this occurs at $\varepsilon_n$ approximately 3~MeV. A fit to the
surrogate decay data in this region will be less sensitive to the
predicted $J\Pi$ population than at lower energies. In addition, the
sensitivity studies for the calculated \ng\ cross section, presented in
\sec{sec:sensitivity}, showed that modeling errors in the s-wave average
radiative width $\left\langle \Gamma_\gamma \right\rangle_0$, or in the
level spacing $D_0$, change primarily the magnitude of the calculated
cross section, but do not affect the energy dependence of the cross
section very much. Therefore, one may try to normalize the decay model
in the \we\ region, where the modeled quantities are not very sensitive
to the $J\Pi$ population. The extracted scaling factor is then used in
the final calculation of the desired cross section at small
energies. The normalization simply becomes an energy- and
$J\Pi$-independent factor, $\eta$.

\begin{figure*}[phbt]
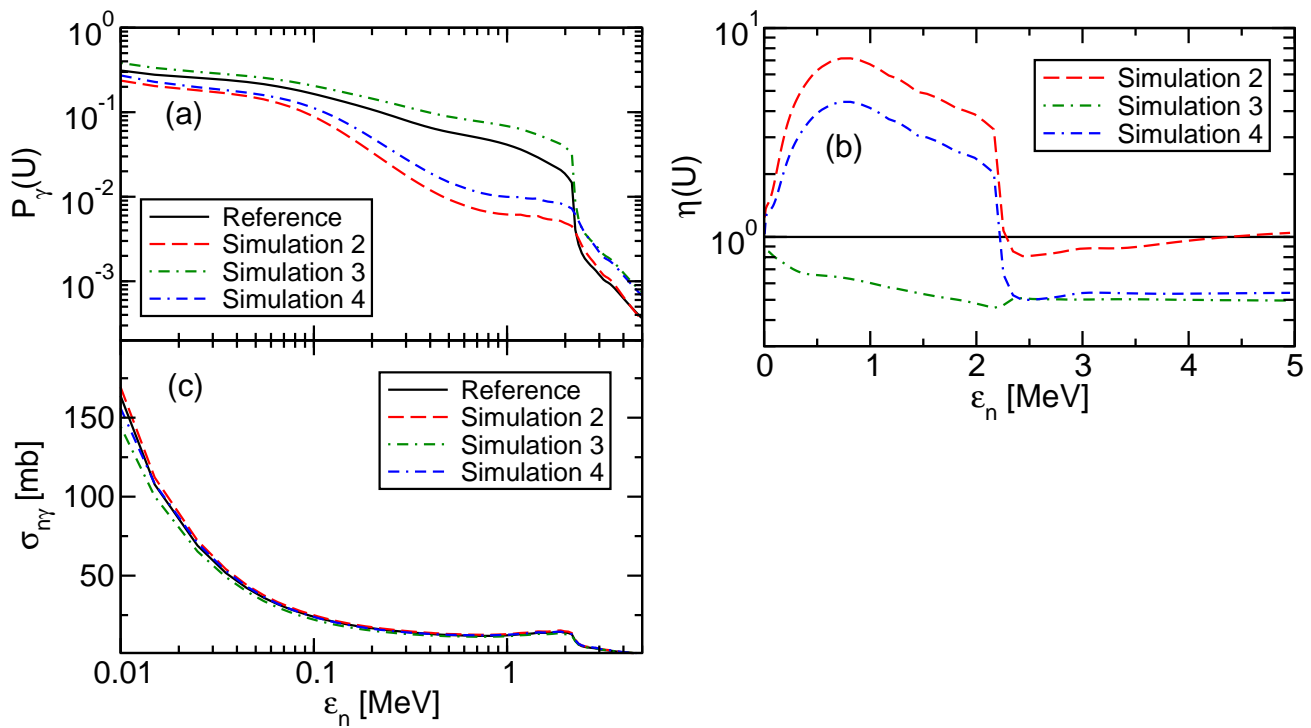

  \begin{minipage}{0.96\textwidth}
    \begin{minipage}[t]{0.49\textwidth}
      \centering
      \mbox{}\\
      \includegraphics*[width=\textwidth]
		       {fig9a_Zr91-GS-ng_surranalysis_approach3.eps}\\
      \mbox{}
    \end{minipage}
  \hfill
    \begin{minipage}[t]{0.49\textwidth}
      \centering
      \mbox{}\\
      \includegraphics*[width=\textwidth]
		       {fig9b_Zr91-GS-ng_surranalysis_approach3eta.eps}\\
      \mbox{}
    \end{minipage}
  \end{minipage}
  \caption{\co\ (a) Gamma-decay probabilities for different
  simulations. The solid line corresponds to the reference model whereas
  the other three are based on different combinations of the assumed
  $J\Pi$ population and decay model (see Table~\ref{tab:sim}).  (b) The
  normalization functions $\eta(U)$. Their value at
  $\varepsilon_n=3$~MeV is used in the final cross section calculation.
  Note that $U = \varepsilon_n + S_n$.
  (c) The extracted cross sections, for the respective simulations,
  obtained by employing the analysis approach outlined in
  \sec{sec:sim3}. The solid line is the reference cross
  section. % 
  \label{fig:sim3}}
\end{figure*}

We examine the outcome of this approach for three different
simulations. As before, we employ the decay probability
$P^\rmref_\gamma(U)$ to represent the measured quantity in our simulated
surrogate experiment. It is shown as a solid line in
\fig{fig:sim3}(a). The dashed line corresponds to the modeled decay
probability obtained from Simulation 2, see Table~\ref{tab:sim}, using
the schematic distribution B (representing a large error compared to the
reference ``true'' distribution), combined with Decay model 2 (yielding
gamma-decay branching ratios that are overestimated by $\sim$20\%). The
dash-dotted line is the modeled decay probability of Simulation 3, in
which the schematic distribution $F^{A+\delta}(J,\Pi)$ is used in
combination with Decay model 1. In this case, the modeled $J\Pi$
population is very close to the reference distribution, but the decay
model is one that produces gamma-decay branching ratios that are known
to be a factor two too large. The dash-dash-dotted line corresponds to
Simulation 4 in which we still use Decay model 1, but the population of
the intermediate state is described by the schematic distribution
B. From \fig{fig:sim3}(a) we observe that the calculated gamma-decay
probabilities differ by up to an order of magnitude in the most relevant
energy regime.

The next step in the analysis procedure is to construct the
normalization functions $\eta(U)$ as described in \eq{eq:fitetaU}. The
resulting functions are shown in \fig{fig:sim3}(b). In this analysis
approach, we focus on the high-energy region in which the normalization
functions have become almost energy-independent. This is the region in
which the \we\ limit is approximately applicable. Still, we want to
extract the normalization at an energy which is not too far away from
the region of interest. In this particular case we use the energy
$\varepsilon_n=3$~MeV, and for our three different simulations we obtain
the normalization factors $\eta = $ 0.88, 0.50, and 0.54,
respectively. The final step is simply to apply this constant
renormalization to our modeling of the desired reaction in
\eq{eq:sigsim2}. Again, we introduce width fluctuation corrections at
this stage. Note that a different decay model was used in Simulation 2
as compared to Simulations 3 and 4, and that both of them differ from
the reference (``true'') decay model. The final result of this procedure
is shown in \fig{fig:sim3}(c) where the extracted cross sections are
compared to the reference result represented by the solid line. The
quality of the extracted cross sections is remarkable considering the
very different initial choices of decay models and $J\Pi$
populations. The MACS at 30~keV for the three simulations are 65, 58,
and 62~mb, respectively, to be compared to the 62~mb for the reference
cross section.

The underlying reason for the success of this approach is the direct
proportionality between variations of the level density formula, or
gamma-strength function, and the corresponding effect on the gamma-decay
branching ratios as demonstrated, e.g., in \fig{fig:ngcs_sens}. In the
\we\ limit this proportionality leads to a universal and
energy-independent scaling of all $J\Pi$ components. This observation
promises to be very useful for surrogate experiments in which data can
be obtained for equivalent neutron energies at which the \we\ limit is
applicable. However, as we have seen in the \zrtwo\ example, see
\fig{fig:sim3}(a), the surrogate gamma-decay probabilities are quite
small for energies above $\varepsilon_n \sim 1$~MeV, and consequently the measurement
is challenging. The conclusion is that, given good-quality surrogate
data in this energy region, the normalization approach outlined above
offers an almost model-independent way of extracting the low-energy \ng\
cross section for spherical and near-spherical nuclei.
%
%**************************************************************************
\section{\label{sec:conc}Conclusions and recommendations}
We have examined the feasibility of using the surrogate method to
extract low-energy \ng\ cross sections for spherical and near-spherical
nuclei in the mass 90-100 region. In particular, several Zr isotopes
were studied and the \zrone\ng\zrtwo\ reaction was used as a test
case. Our study was performed by carrying out simulations to explore
three different approaches to utilize surrogate reaction data. The
sensitivity of the extracted cross sections to uncertainties in the
theoretical modeling was investigated for the different approaches and
their performance was assessed by comparing with a predetermined
benchmark.

One of the main results of the paper is the demonstration of the large
sensitivity of the gamma-decay branching ratios, and consequently the
surrogate-reaction gamma-decay probability, to the $J\Pi$ population of
the intermediate \cn, and the effect that this sensitivity has on the
extracted cross section. It was found that an approach in which the
$J\Pi$ population mismatch was not taken into account, gave an extracted
cross section that deviated from the benchmark result by an order of
magnitude. On the other hand, full modeling of the population of the
\cn\ is very difficult as it involves the description of direct
reactions to continuum states and their subsequent equilibration into a
compound-nuclear state. The third and final analysis approach proposed
in this paper promises to be the most viable one. It begins with a
careful modeling of the decay, preferably using regional systematics to
constrain unknown nuclear properties. Surrogate decay data collected at
slightly higher energies are then utilized to normalize the modeled
gamma-decay branching ratios so that the desired cross section can be
extracted.

It needs to be emphasized that these findings, and the sensitivities
that were studied, apply to the analysis of our simulated surrogate
experiment. In actual experiments, additional uncertainties are
introduced, e.g., the identification of the final state, and the finite
energy and angular resolutions. Furthermore, a prerequisite for our
simulations of surrogate reactions has been the population of a fully
equilibrated compound-nuclear intermediate state. The question how this
state is formed, through the population of highly-excited states in the
direct surrogate reaction followed by multi-step equilibration
processes, deserves to be studied in much greater detail. In particular,
the probability for pre-equilibrium emission of particles should be
investigated for different surrogate reactions.

The theoretical modeling in this paper was performed using \hf\
statistical-reaction theory. The approaches for normalizing the decay
model using surrogate-reaction experimental data is based on the use of
traditional decay models. Furthermore, these decay models were applied
to describe the decay from all relevant $J\Pi$ states in the entire
energy range of interest.  This implies that the arguments presented
here are not applicable to regions of nuclei that are too far away from
the valley of stability, where the use of established level-density
formulae and parameterized gamma-strength functions has to be
questioned. The same observation applies to the optical potentials that
were used in this work to calculate particle transmission
coefficients. The normalization procedures outlined in
Sec.~\ref{sec:analysis} rests on the accuracy of the optical model that
is used to compute the absorption cross section. Finally, we stress that
width fluctuation corrections were always applied in the final modeling
part of the analysis. Those corrections cannot be obtained from the
surrogate experimental data.

With the insights gained from this study we can give a few general
recommendations for future surrogate experiments: (1) Good particle
identification is very important since one needs the absolute
normalization to get the decay probabilities with good accuracy. We note
that the gamma-decay probability is expected to decrease dramatically in
magnitude and one must be able to measure them up to a few MeV above the
neutron separation energy. (2) Consequently one also needs to be able to
efficiently identify the desired decay channel. For gamma decay this
might require the existence of a very strong collector $2^+ \rightarrow
0^+$ transition to tag the occurrence of a gamma cascade. This
requirement would limit the method to cases where the intermediate
nucleus is an even-even isotope. However, other options for tagging the
gamma cascade are being considered~\cite{ber06:pc}. (3) The energy
resolution is less of an issue if the approach of renormalizing the
decay model at high energies is used. In this procedure the energy
dependence is obtained from the statistical-reaction modeling. Still, an
energy resolution that is better than 100~keV should be desirable. (4)
Additional experimental information, such as relative intensities of the
gammas de-exciting low-lying levels of different spin and parity, can
help to gain insight into the $J\Pi$ population of the \cn.  However,
one should remember that the observed gamma intensities depend on the
properties of the gamma cascade that proceeds through the
quasi-continuum as well as on the initial $J\Pi$ distribution, and so it
is important that the properties of this cascade be accurately
modeled. (5) Measuring the decay at different emission angles of the
ejectile in the initial direct reaction should provide an experimental
handle to vary the $J\Pi$ population. However, it remains to be seen
from direct-reaction modeling for each particular case how large this
change can actually be. (6) Finally, we recommend that benchmark
experiments, in which the surrogate data is used to extract a known
cross section, are carried out since this will provide very valuable
insights into the issues discussed in this paper.

Finally, we note that applications involving spherical nuclei probably
constitute one of the most challenging applications of the surrogate
method. This statement applies, in particular, to isotopes near magic
shells such as the Zr isotopes studied in the present work.  In
addition, low-energy \ng\ reactions provide the most difficult reaction
with regards to the angular momentum mismatch. In contrast, for deformed
nuclei there is usually a larger number of open decay channels and that
reduces the sensitivity of the gamma-decay branching ratios to the
initial $J\Pi$ population. In particular, the \we\ limit might be
reached already at relatively low excitation energies above the neutron
separation energy.
%
%**************************************************************************
\begin{acknowledgments}
This work was partly performed under the auspices of the U.S. Department
of Energy (DOE) by the University of California, Lawrence Livermore
National Laboratory (LLNL) under contract No.~W-7405-Eng-48. Support
from the Laboratory Directed Research and Development Program at LLNL,
contract No.~04--ERD--057, is acknowledged.
\end{acknowledgments}
%**************************************************************************

%**************************************************************************
\end{document}